\documentclass[twocolumn,prb,superscriptaddress,showpacs]{revtex4}

\usepackage{graphicx}
\usepackage{amsmath,graphicx,dcolumn}
\usepackage[usenames]{color}

\begin{document}

\title{Local structure and site occupancy of Cd and Hg substitutions\\
 in Ce$T$In$_{5}$ ($T$=Co, Rh, Ir)}

\author{C. H. Booth}
\affiliation{Chemical Sciences Division,
Lawrence Berkeley National Laboratory,
Berkeley, California 94720, USA}
\author{E. D. Bauer}
\affiliation{Materials Physics and Applications Division, Los Alamos National 
Laboratory, Los Alamos, New Mexico 87545, USA}
\author{A. D. Bianchi}
\affiliation{Physics Department, University of California, Irvine, CA 92697, USA}
\affiliation{Département de Physique et RQMP, Université de Montréal, Montréal, QC H3C 3J7, Canada}
\author{F. Ronning}
\affiliation{Materials Physics and Applications Division, Los Alamos National 
Laboratory, Los Alamos, New Mexico 87545, USA}
\author{J. D. Thompson}
\affiliation{Materials Physics and Applications Division, Los Alamos National 
Laboratory, Los Alamos, New Mexico 87545, USA}
\author{J. L. Sarrao}
\affiliation{Materials Physics and Applications Division, 
Los Alamos National Laboratory, 
Los Alamos, New Mexico 87545, USA}
\author{Jung Young Cho}
\affiliation{Department of Chemistry, Louisiana State University,
Baton Rouge, LA 70803-1804, USA }
\author{Julia Y. Chan}
\affiliation{Department of Chemistry, Louisiana State University,
Baton Rouge, LA 70803-1804, USA }
\author{C. Capan}
\affiliation{Physics Department, University of California, Irvine, CA 92697, USA}
\author{Z. Fisk}
\affiliation{Physics Department, University of California, Irvine, CA 92697, USA}

\date{submitted to Phys. Rev. B Feb. 5, 2009, accepted March 25, 2009}


\begin{abstract}
The Ce$T$In$_5$ superconductors ($T$=Co, Rh, or Ir) have generated great 
interest due to their relatively high transition temperatures, non-Fermi liquid
behavior, and their proximity to antiferromagnetic order
and quantum critical points. In contrast to small changes with the $T$-species,
electron doping in Ce$T$(In$_{1-x}M_x)_5$ with $M$=Sn and hole doping with Cd 
or Hg have a dramatic effect on the electronic properties at very low 
concentrations.  The present work reports local structure measurements using 
the extended x-ray absorption fine-structure (EXAFS) technique that address
the substituent atom distribution as a function of $T$, $M$, and $x$, in the 
vicinity of the superconducting phase. Together with previous measurements for 
$M$=Sn, the proportion of the $M$ atom residing on the In(1) site, 
$f_\textrm{In(1)}$, increases in the order $M$=Cd, Sn, and Hg, ranging from 
about 40\% to 70\%, showing a strong preference for each of these substituents 
to occupy the In(1) site (random occupation = 20\%).  In addition, 
$f_\textrm{In(1)}$ ranges from 70\% to 100\% for $M$=Hg in the order $T$=Co, 
Rh, and Ir.  These fractions track the changes in the atomic radii of the 
various species, and help explain the sharp dependence of $T_c$ on substituting 
into the In site. However, it is difficult to reconcile the small concentrations
of $M$ with the dramatic changes in the ground state in the hole-doped 
materials with only an impurity scattering model.  These results therefore
indicate that while such substitutions have interesting local atomic structures 
with important electronic and magnetic consequences, other local changes in the
electronic and magnetic structure are equally important in determining the bulk 
properties of these materials.
\end{abstract}



\maketitle

\section{Introduction}

\begin{figure}[b]
\includegraphics[width=1.8in]{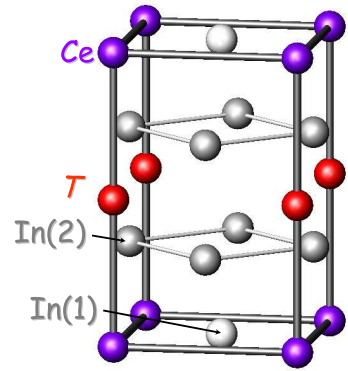}
\caption{(Color online) The tetragonal unit cell of the Ce-115s.
}
\label{xtal}
\end{figure}

The rich variety of novel strongly-correlated electron phenomena observed in the 
family of Ce$T$In$_5$ ($T$=Group VIII transition metal) heavy-fermion 
compounds,\cite{Thompson03} such as the coexistence of unconventional superconductivity 
and magnetism under pressure\cite{Hegger00,Park06} or through chemical 
substitution,\cite{Pagliuso01,Zapf01} and magnetic field-induced magnetism within the 
superconducting state,\cite{Bianchi03b,Young07,Mitrovic07} has invigorated interest in 
understanding the interplay of superconductivity and magnetism in strongly-correlated 
materials.  The Ce$T$In$_5$ family (generically referred to as the ``115s'') is ideally 
suited to explore this 
interplay as the energy scales of these two ground states are easily tuned with 
modest pressures or magnetic fields. Recent work has focused on the effects of
substitutions onto the In sites (Fig. \ref{xtal}), effectively either electron doping with 
Sn\cite{Bauer06} or hole doping with Cd\cite{Pham06} or Hg.\cite{Bauer08} 
Previous local structure studies of the atomic environment around the Sn atoms 
using the extended x-ray absorption fine-structure (EXAFS) technique found that 
Sn atoms preferentially reside on the In(1) site, helping explain 
the sharp dependence of the superconducting (SC) transition temperature, $T_c$, on 
the Sn concentration and further supporting the notion of quasi-two-dimensional
superconductivity confined primarily to the Ce-In(1) planes.\cite{Daniel05b} 
Subsequent studies have shown that hole doping produces even more dramatic effects, 
including accessing the antiferromagnetic (AFM) phase 
and exhibiting reversible behavior under applied pressure.\cite{Pham06,Bauer08}
It is therefore vital to determine the distribution of Cd and Hg on the
In sites in these materials to properly assess the role that impurity
scattering plays in the properties of the hole-doped 115 materials.
The present study extends the previous study on CeCo(In$_{1-x}$Sn$_x$)$_5$ and
 determines these distributions using the EXAFS technique as a
function of the species of $M$ in CeCo(In$_{1-x}M_x$)$_5$ with $M$ = Cd and Hg,
and as a function of $T$ in Ce$T$(In$_{1-x}$Hg$_x$)$_5$ with $T$ = Co, Rh, and Ir.

\begin{figure}[t]
\includegraphics[width=3.5in]{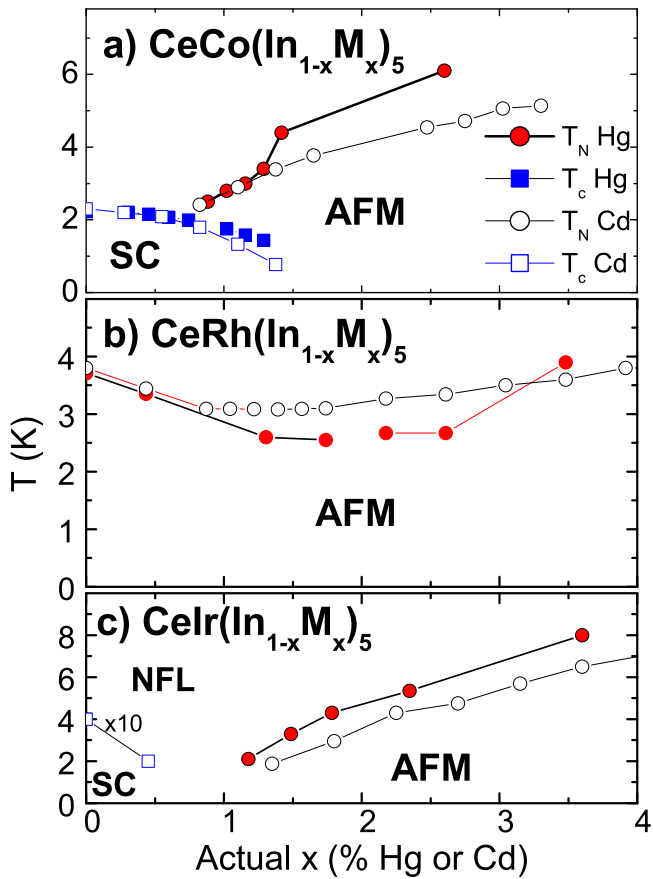}
\caption{(Color online) Phase diagrams of Cd- and Hg-substituted Ce$T$In$_5$.
The reported substituent concentrations of Hg-substituted samples are 
estimated from microprobe measurements of the Cd-substituted samples. See Sec. \ref{Experimental} for details.
}
\label{phase}
\end{figure}

The substitution of Cd or Hg for In at the percent level in Ce$T$In$_5$ has revealed 
a simple way to continuously tune between SC and AFM
order (Fig. \ref{phase}) while introducing minimal structural disorder.  
In particular, $T_c$ remains nearly constant with increasing Cd substitution up to 
$x$=0.5\% from 2.3~K in pure CeCoIn$_5$. Superconductivity coexists with 
long-range AFM order 
for 0.5\%\ $< x \leq 1.25$\%, after which point only AFM is observed.\cite{Pham06} 
(Note that concentrations as measured by microprobe measurements of Cd and Hg are
reported
throughout this article, which are very close to 10\% of the nominal concentration reported 
previously.\cite{Pham06})  The entropy balance between these ground 
states along with the observation of coupled SC and AFM order parameters by neutron 
diffraction, implies that the same electronic degrees of freedom determine the nature 
of the ground state in these materials.\cite{Nicklas07}  
The application of pressure to 
CeCoIn$_5$:Cd reverses the evolution of the ground state with Cd substitution and 
also mimics the pressure-induced behavior of CeRhIn$_5$,\cite{Park06} in which 
AFM order is suppressed from its ambient pressure value of $T_\textrm{N}=3.8$~K to zero 
temperature by $\sim$2.3 GPa, and coexists with superconductivity in an intermediate 
pressure range between 0.5 GPa and 1.7 GPa. However, for the small Cd concentrations that
induce these changes, there is little detectable change in the average structure.\cite{Pham06}
These results suggest that it is the 
slight hole-doping of CeCoIn$_5$ with Cd that tunes the electronic structure sufficiently to 
induce magnetism, rather than chemical pressure or disorder effects.  

This behavior contrasts with that achieved by electron doping 
with Sn into Ce$T$In$_5$, which 
uniformly suppresses AFM order in CeRhIn$_5$ by 7\% Sn for In without inducing 
superconductivity,\cite{Bauer06} and completely suppresses superconductivity in 
CeCoIn$_5$ at 3-4\% Sn with no sign of antiferromagnetic order.\cite{Bauer05,Bauer06a}
This behavior is more congruous with that achieved by substituting with La on the Ce 
site,\cite{Petrovic02} especially when considering the propensity of Sn to rest on the
in-plane In(1) site.\cite{Daniel05b} Even so, the reduction of $T_c$ with the Sn 
In(1)-site occupancy remains sharper compared to La substitutions,
providing further evidence that slight changes in electronic structure dominate the 
underlying physics in the substituted Ce$T$In$_5$ materials. Although 
Abrikosov-Gorkov-like\cite{Abrikosov61,Muller71,Daniel05b,Paglione07} pair-breaking undoubtedly plays some role, exactly
how such minute quantities of 
these particular substituent atoms are able to tip the delicate balance between the nearly 
degenerate SC and AFM ground states in Ce$T$In$_5$, where 
substitution on the transition metal site requires of order 30\% to induce similar 
changes, is an important, yet poorly understood issue in the interplay between these 
two phenomena. 

Here, the local structure around Cd and Hg in Ce$T$In$_5$ using the 
EXAFS technique is reported to determine how the local environment affects the ensuing magnetism 
and superconductivity. The EXAFS technique, while only having a range of about
6~\AA, provides a particularly powerful way
of determining the local atomic environment around the substituent atoms, because a
specific core-electron x-ray absorption process is chosen. Therefore, even though
very little Cd or Hg exist in these materials, only scattering paths involving
Cd or Hg contribute to the EXAFS signal. The main structural difference for 
differentiating between the In(1) and In(2) sites is the nearest-neighbor 
In(2)-$T$ distance at about 2.8~\AA, since the nearest neighbors to the In(1) site 
are Ce and In(2) at about 3.3~\AA. Other differences in the local structures 
around the In(1) and In(2) sites also help determine the fraction of the substituent 
atoms on the In sites. In addition, EXAFS is useful for determining distortions from the
average crystal structure, which may also be important in determining the effects of
substitutions onto the In sites.

The rest of this article is organized as follows: experimental
methods and data fitting techniques are described in Sec. \ref{Experimental}, while the
details of the results of the fits are in Sec. \ref{Results}. These results are
related to various parameters in Sec. \ref{Discussion}, such as $T_c$, the various atomic radii and the substituent In(1)-site occupancy. Finally, the main
conclusions of this research are summarized in Sec. \ref{Conclusions}.

\begin{figure}[t]
\includegraphics[width=3.5in,trim=0 0 0 0]{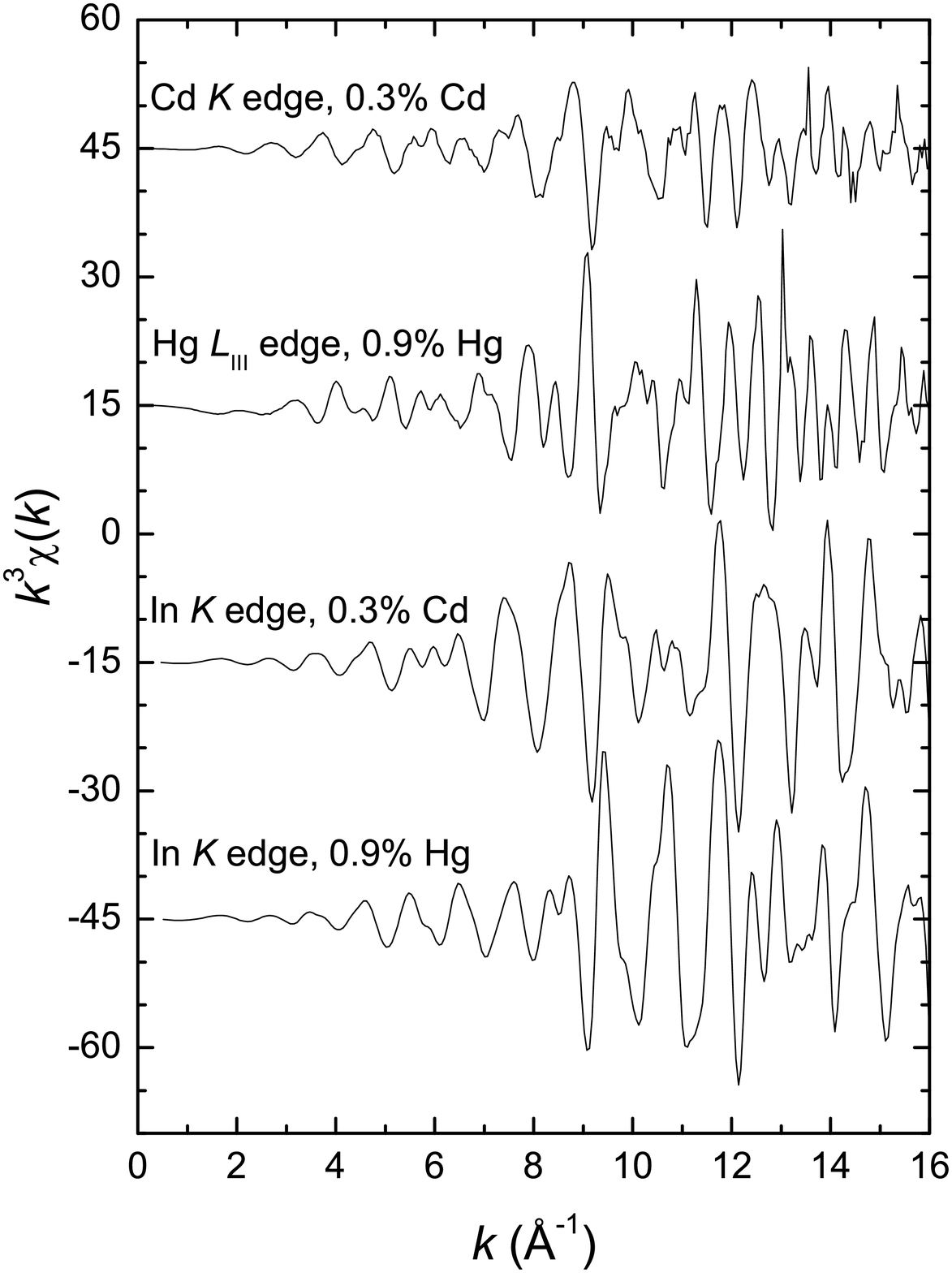}
\caption{Examples of $k$-space data at 30 K for representative samples
at all four measured edges. These data are from averages of between 3 and 6 scans,
each measured over about a half of an hour.
}
\label{ks_fig}
\end{figure}

\section{Experimental Methods}
\label{Experimental}
Samples were synthesized as described in Refs. \onlinecite{Pham06,Bauer08}.  
X-ray diffraction measurements on the Cd-substituted samples indicate
contraction of both the $a$ and the $c$ lattice parameters of about 0.005 \AA{}
in the vicinity of the critical concentration $x_c$ where the samples cease to
be superconducting (Fig. \ref{phase}).  Measurements on the Sn- and 
Hg-substituted samples, however, have not been able to identify any clear trend 
in the lattice parameters with concentration.
Microprobe analysis of
CeCo(In$_{1-x}$Cd$_x$)$_5$ yielded an actual/nominal Cd concentration ratio $x_{act}/x_{nom} = 0.11$, close to
the value of 0.10 reported in Ref. \onlinecite{Pham06}.  A similar analysis\cite{Bauer09} yielded an actual/nominal Hg concentration ratio of  0.16, 0.17, 0.18 for $T$= Co, Rh, Ir, respectively.  Lacking microprobe
data for CeRh(In$_{1-x}$Cd$_x$)$_5$ and CeIr(In$_{1-x}$Cd$_x$)$_5$, the same  actual/nominal Cd concentration ratio was used for a relative comparison to the Hg-doped CeRhIn$_5$ and CeIrIn$_5$ temperature-composition phase diagrams, given that Cd is isoelectronic with Hg.   

The following
samples were measured with the EXAFS technique for this study, although not all
data are explicitly reported for the sake of brevity: CeCo(In$_{1-x}$Cd$_x$)$_5$ with
$x$=0.003, 0.005, 0.011 and 0.18; CeCo(In$_{1-x}$Hg$_x$)$_5$ with $x$=0.007, 0.012,
and 0.014; CeRh(In$_{1-x}$Hg$_x$)$_5$ with $x$=0.009, 0.026, and 0.035; and 
CeIr(In$_{1-x}$Cd$_x$)$_5$ with $x$=0.009, 0.018, and 0.036.

X-ray absorption data were collected at Beamline 11-2 of the Stanford Synchrotron
Radiation Lightsource using half-tuned Si(220) monochromator crystals on the
unfocused beam. The samples were prepared for these absorption measurements by
grinding them in a mortar and pestle under
acetone, with the resulting powder passed through a 32 $\mu$m sieve. This
powder was brushed onto adhesive tape, which was then cut into strips and
stacked, either in sufficient quantity to have reasonable fluorescence data
from the Cd $K$ and Hg $L_\textrm{III}$ edges, or to obtain a change in the absorption
across the In $K$ edge of about 0.8 absorption lengths. The samples were placed in a liquid helium
flow cryostat at 30~K. Data at the Cd $K$ or Hg $L_\textrm{III}$ edges were collected in
fluorescence mode and corrected for the dead time of the 32-element Ge detector. 

Data reduction and fitting were performed using the RSXAP 
package\cite{Hayes82,RSXAP} with scattering functions generated by the 
FEFF7 code.\cite{FEFF7} 
In particular, data collected in transmission mode must be treated differently
than data collected in fluorescence mode. For the In $K$-edge data collected
in transmission mode,
the absorption contribution from the desired core excitation, 
$\mu_a$, was
isolated from the total absorption, by subtracting 
the contribution from other absorption processes, as 
determined from a fit to the pre-edge data and forcing the remaining
absorption to follow a Victoreen formula.\cite{Li95b} 
The embedded atom absorption $\mu_0$
was generated by fitting a 7-knot cubic spline function through the data above
the main absorption edge.  The EXAFS function was then calculated using 
$\chi(k)=[\mu_a(k)-\mu_0(k)]/\mu_0(k)$, where $k=[2m_e(E-E_0)/\hbar^2]^{1/2}$,
$E$ is the incident photon energy, and $E_0$ is the photoelectron threshold
energy as determined by the position of the half-height of the edge. Fluorescence
data are treated similarly, but there are two important differences. 
First, the absorption processes from channels other than the desired excitation
are already discriminated against by the energy-sensitive Ge detector, apart from
much smaller corrections due to roughly constant background processes and Compton
scattering of the direct beam into the energy window for the desired Cd $K_\alpha$
or Hg $L_\alpha$ fluorescence lines. For each absorption process, a
fluorescence photon is generated, so overall changes in the fluorescence above the
absorption edge already should include the overall decrease in the absorption
described by the Victoreen formula. For these reasons, a different 
pre-edge background, $\mu_\textrm{pre}$ is applied that only tries to isolate the
desired fluorescence line. Second, self-absorption processes can play an important 
role, and are, in fact, the main factor in overall increases or decreases in the 
observed fluorescence.\cite{Goulon82,Troger92,Booth05a} A self-absorption 
correction\cite{Booth05a} was applied, but was typically less than 2\%.  Examples
of these data are shown in Fig. \ref{ks_fig} as an illustration of the quality of the data.
Note that all data  were collected to 16.0 \AA,$^{-1}$ except the
Hg $L_\textrm{III}$-edge (12284 eV) data for CeIr(In$_{1-x}$Hg$_x$)$_5$, which was 
limited to 11.5 \AA$^{-1}$ by the Ir $L_\textrm{II}$ edge (12824 eV).

The 115 local structure around Ce and $T$-site atoms is relatively simple, with
well separated scattering shells. The local structure around In is much more
complicated, owing both to the two In sites and to strong overlap between
In(1)-Ce, In(1)-In(2), In(2)-In(2), In(2)-In(1), and In(2)-Ce near neighbors,
which are all near 3.3 \AA. Although 
substituent atoms should appear in the backscattering [for instance, the In(1)-In(2) 
peak will overlap an In(1)-Hg(2) peak], such scattering shells have an insignificant
contribution at the measured substituent concentrations. Such peaks are, in any case, included 
in the fitting model.
In all, there are 20 single-scattering paths up to 
5 \AA. The bond lengths in the fitting model are therefore tightly constrained to the 
nominal 115 
structure, equivalent to only allowing the variation of the $a$ and $c$ lattice parameters,
the position $z$ along the $c$ axis of the In(2) sites, and, when the data allows,
two additional atom-pair distances. Only the Cd-edge data and the Hg-edge data on the 
substituted  CeIrIn$_5$ sample required tightening these constraints. 
In addition,
many of the mean-squared displacements of the pair distances, $\sigma^2$'s,
are also constrained together. The number of neighbors per absorbing atom,
$N$, are constrained to the nominal
values, allowing both for an overall scale factor in the fit, $S_0^2$, the
fraction of the absorbing atom on the In(1) site, $f_\textrm{In(1)}$, and the
$x$ value for the substituent concentration, only the latter of which is held fixed. 
Discrepancies between the
actual structure and a fitting model of this type will manifest as enhanced 
values of the $\sigma^2$ parameters. 
Note that, generally, only scattering paths with independent bond lengths in 
this model are reported in the 
tables for simplicity. The remaining independent parameters that are not
explicitly reported are $\sigma^2$ parameters. 
However, all of these
fall within reasonable limits, never exceeding about 0.006 \AA$^2$.

Reported errors are determined using a Monte Carlo method\cite{Lawrence01} that
do not properly account for systematic errors. The possible magnitude of systematic 
errors are discussed in Sec. \ref{Results} below.

\section{Results}
\label{Results}

\begin{figure}[t]
\includegraphics[width=3.5in]{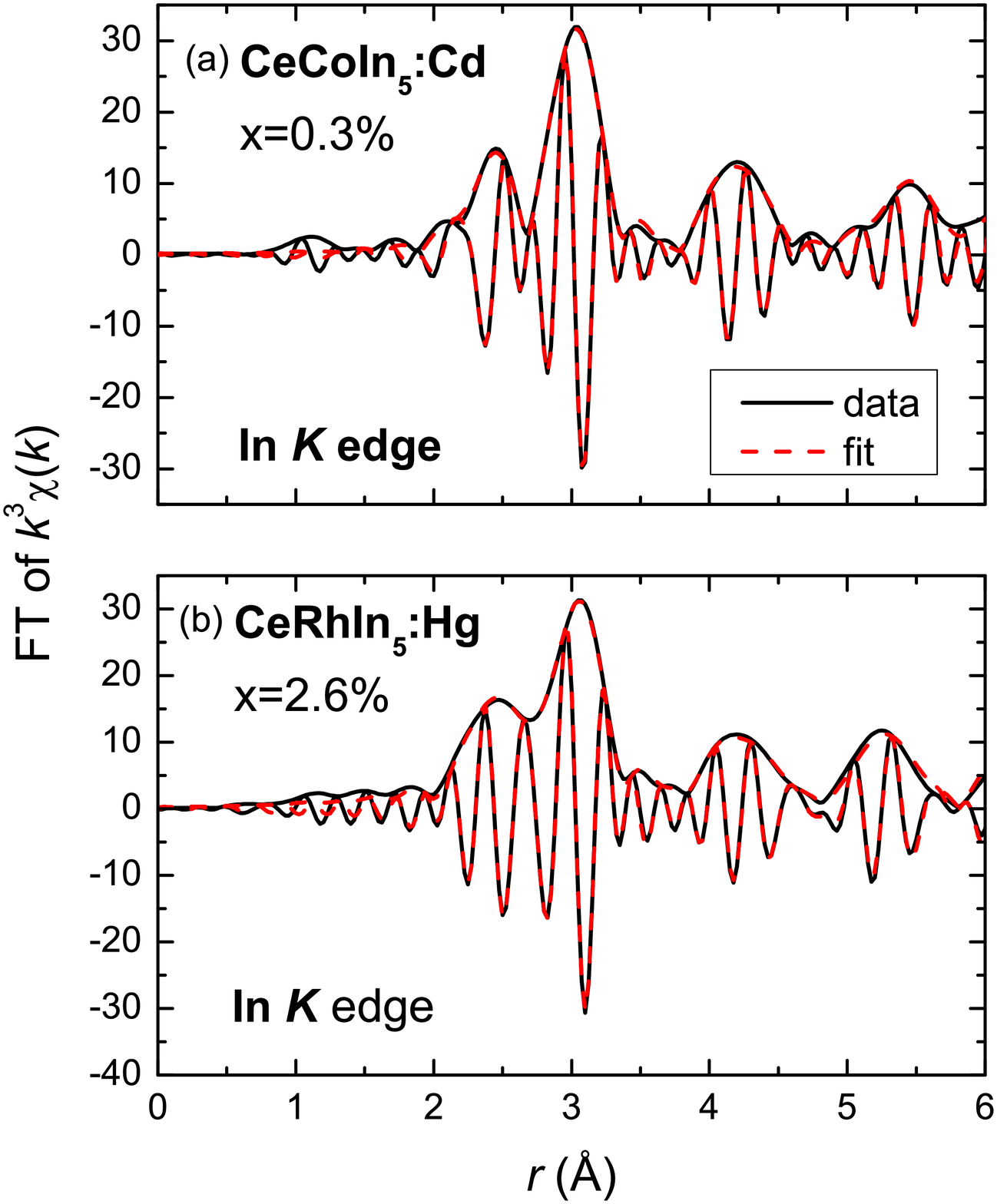}
\caption{(Color online) Representative fits to Fourier transform (FT) of $k^3(\chi(k)$ 
In $K$-edge data.  All transforms are between 2.5-16.0~\AA$^{-1}$ after Gaussian 
narrowing by 0.3~\AA,$^{-1}$ unless otherwise noted.  The outer envelope is the 
modulus, and the oscillating line is the real part of the complex transform. The 
difference in Co and Rh backscattering is demonstrated by the large difference in the 
first peak near 2.5 \AA, due primarily to (a) In(2)-Co or (b) In(2)-Rh scattering paths.
}
\label{inrs_fig}
\end{figure}

Fourier transforms (FTs) of the $k^3\chi(k)$ data from the In $K$ edge are shown in
Fig. \ref{inrs_fig}, which demonstrate several of the important features of the other
transforms discussed in 
this article. The largest peak is due to several overlapping In-In and In-Ce
pairs, as discussed above and indicated in the Tables. The peak position $r$
is shifted from the actual pair distance $R$ due to a phase shift of the
photoelectron that occurs both as the electron leaves and returns to the 
absorbing atom and at the backscattering atom. This shift is well reproduced by 
FEFF7, allowing accurate bond lengths to be determined from the fits.\cite{Li95b} The real 
part of the transform, which is shown as the oscillating line between the
modulus envelope in the FT figures, gives an indication of this phase shift as
a function of the backscattering atomic species. In particular, Rh is both
a stronger back-scatterer than Co, and has a much larger phase shift. Therefore, 
the In-Rh peak near $r\sim2.5$ \AA{} is larger, but nearly out of phase
with the In-Co peak. 
Transforms of In $K$-edge data at different $M$ concentrations (not shown) change 
very little, indicating the small
effect each substituent has on the average crystal structure.

\begin{table}
\caption{Fit results for In $K$-edge data at 30 K on 
CeRh(In$_{0.991}$Hg$_{0.009}$)$_5$. All scattering paths are included within the
fitting range, but only those single-scattering paths with independent pair distances are
reported here. All other path distances are constrained to these paths. 
Fit range is between 2.2 and 6.1 \AA. The $k^3$-weighted data are transformed between 
2.5-16.0~\AA$^{-1}$ after Gaussian narrowing by 0.3~\AA.$^{-1}$
These fits have about 20 degrees of freedom.\cite{Stern93}
Reported errors in parentheses are from a 
Monte Carlo method and represent the random error associated with the fit. 
Systematic errors are discussed in the text. See text for further details.
Note that the In $K$-edge fit results are primarily used to test the efficacy 
of the fitting model for determining the fraction of the absorbing atom on the 
In(1) site, $f_\textrm{In(1)}$, which in nominally 0.2 for indium absorbers.}
\begin{ruledtabular}
\begin{tabular}{lcccc}
\\
& $R_\textrm{diff}$\footnotemark[1] (\AA) & $N$ & $\sigma^{2}$(\AA$^2$) & $R$(\AA) \\
\colrule
In(1)-Ce & 3.2923
&  0.47&  0.0017(7) &  3.270(4) \\
In(1)-In(2) & 3.2775
&  0.94&  0.0027(5) &   3.23(6) \\
In(2)-Rh & 2.7500
&  1.76&  0.0023(3) &  2.736(2) \\
In(2)-Ce & 3.2775
&  1.76& 0.0017\footnotemark[2]&   3.27(1) \\
In(2)-In(2) & 4.6142
&  0.87&   0.001(1) &   4.64(1) \\
\\
$\Delta E_0$  &  -5.6(6)  && \\
$S_0^2$ &   0.94(6)  && \\
$f_\textrm{In(1)}$ &   0.12(3)  && \\
$R(\%)$  &  6.13 && \\
\end{tabular}
\end{ruledtabular}
\footnotetext[1]{From Ref. \onlinecite{Moshopoulou02} for CeRhIn$_5$}
\footnotetext[2]{Constrained to In(1)-Ce}
\label{in_tbl}
\end{table}

A fit of this structure to the In $K$-edge data from one of the samples is 
shown in Fig. \ref{inrs_fig} and the results are given in Table \ref{in_tbl} for CeRhIn($_{1-x}$Hg$_x$)$_5$. 
The primary purpose of such fits is to demonstrate the efficacy 
of the fitting model, and therefore the final results are compared
to the nominal crystal structure to help quantify any systematic errors.
To this end, the measured pair distances are all close to those measured by
diffraction, although outside the estimated errors. Considering that only 5 fit
parameters describe all the bond lengths up to 5 \AA, the systematic errors in the 
pair distances are expected to be within about 0.02 \AA,\cite{Li95b} as observed. 
The $\sigma^2$ parameters are all small, as expected for a well-ordered crystal 
lattice. Of particular interest is the fraction of In atoms
on the In(1) site, which is nominally 0.2. Within this fitting model,
$f_\textrm{In(1)}=0.12(5)$. The fits to the In $K$-edge data from all the samples
give similar results, so systematic errors in $f_\textrm{In(1)}$ are expected
to be better than 0.1. However, this error may be smaller when a higher fraction
of a particular substituent species resides on the In(1) site, as determined for all the 
samples discussed below.

\begin{table*}
\caption{Fit results for Cd $K$ data at 30 K on CeCo(In$_{1-x}$Cd$_{x}$)$_5$. 
Fit range is between 2.2 and 5.1 \AA. The $k^3$-weighted data are transformed between 
2.5-16.0~\AA$^{-1}$ after Gaussian narrowing by 0.3~\AA.$^{-1}$
These fits have about 13 degrees of freedom.\cite{Stern93}
See Table \ref{in_tbl} and text for further details.}
\begin{ruledtabular}
\begin{tabular}{lcccccccccc}
&
&\multicolumn{3}{c}{$x$=0.003}
&\multicolumn{3}{c}{$x$=0.005}
&\multicolumn{3}{c}{$x$=0.011}
\\
& $R_\textrm{diff}$\footnotemark[1] (\AA)
& $N$ & $\sigma^{2}$(\AA$^2$) & $R$(\AA) & $N$ & $\sigma^{2}$(\AA$^2$) & $R$(\AA) & $N$ & $\sigma^{2}$(\AA$^2$) & $R$(\AA) \\
\colrule
Cd(1)-Ce & 3.2618
&  1.63&  0.0007(6) &  3.253(7) &  1.68&   0.002(1) &  3.246(6) &  1.88&  0.0013(6) &  3.246(4) \\
Cd(1)-In(2) & 3.2830
&  3.25&  0.0043(7) &  3.152(8) &  3.35&   0.006(1) &  3.154(6) &  3.71&  0.0040(6) &  3.157(3) \\
Cd(2)-Co & 2.7187
&  1.19&  0.0009(6) &  2.738(6) &  1.16&  0.0022(8) &  2.724(6) &  1.06&  0.0004(4) &  2.735(3) \\
Cd(2)-Ce & 3.2830
&  1.19& 0.0007\footnotemark[2]     &  3.152    &  1.16&  0.002\footnotemark[2]     &  3.154    &  1.06& 0.0013\footnotemark[2]     &  3.157 \\
\\
$\Delta E_0$  & &  -3(1)  &  & &  -3(1)  &  & &  -1.9(7)  & \\
$S_0^2$ & &   0.9(1)  &  & &   0.9(1)  &  & &   0.9(1)  & \\
$f_\textrm{In(1)}$ & &   0.41(4)  &  & &   0.42(4)  &  & &   0.47(4)  & \\
$R(\%)$  & & 13.6 &  & & 13.2 &  & &  9.9 & \\
\end{tabular}
\end{ruledtabular}
\footnotetext[1]{From Ref. \onlinecite{Moshopoulou02} for CeCoIn$_5$}
\footnotetext[2]{Constrained to Cd(1)-Ce}
\label{cd_tbl}
\end{table*}

\begin{table*}
\caption{Fit results from Hg $L_\textrm{III}$ edge data on 
Ce$T$(In$_{(1-x)}$Hg$_{x}$)$_5$. Fit range is between 2.2 and 6.1 \AA. 
The $k^3$-weighted data are transformed between 2.5-16.0~\AA$^{-1}$, except 
the $T$=Ir data, which are $k$-weighted and transformed between 
2.5-11.5~\AA.$^{-1}$
All data are Gaussian narrowed by 0.3~\AA$^{-1}$ before transforming.
The degrees of freedom for these fits are about 20 for the $T$=Co and Rh data, and 
about 8 for the $T$=Ir data.\cite{Stern93}
See Table \ref{in_tbl} and text for further details.}
\begin{ruledtabular}
\begin{tabular}{lcccccccccc}
&&\multicolumn{3}{c}{$T$=Co, $x$=0.010}
&\multicolumn{3}{c}{$T$=Rh, $x$=0.026}
&\multicolumn{3}{c}{$T$=Ir, $x$=0.018}
\\
& $R_\textrm{diff}$\footnotemark[1] (\AA)
& $N$ & $\sigma^{2}$(\AA$^2$) & $R$(\AA) & $N$ & $\sigma^{2}$(\AA$^2$) & $R$(\AA) & $N$ & $\sigma^{2}$(\AA$^2$) & $R$(\AA) \\
\colrule
Hg(1)-Ce & 3.2923
&  2.78&  0.0022(7) &  3.247(2) &  3.67&  0.0012(8) &  3.278(2) &  4.00&   0.003(3) &   3.27(1) \\
Hg(1)-In(2) & 3.2775
&  5.55&   0.008(1) &   3.22(2) &  7.32&  0.0022(8) &  3.228(3) &  8.00&   0.005(2) &   3.21(1) \\
Hg(2)-Co & 2.7500
&  0.61&   0.007(4) &   2.76(1) &  0.17&    0.02(1) &  2.7(1) &  0.00& - & -\\
Hg(2)-Ce & 3.2775
&  0.61& 0.0022\footnotemark[2]&    3.2(1) &  0.17& 0.0012\footnotemark[2]&   3.1(1) &  0.00& -& -\\
Hg(2)-In(2) & 4.6142
&  0.30&    0.02(1) &   4.58(7) &  0.08& 0.040&   4.8(2) &  0.00& - &  -\\
\\
$\Delta E_0$  & &  -0.3(6)  &  & &  -2.4(6)  &  & &  -0.9(8)  & \\
$S_0^2$ & &   0.56(7)  &  & &   0.77(6)  &  & &   1.0(1) & \\
$f_\textrm{In(1)}$ & &   0.70(8)  &  & &   0.92(4)  &  & &   1.00(4)  & \\
$R(\%)$  & &  9.0 &  & & 12.0 &  & & 27.0 & \\
\end{tabular}
\end{ruledtabular}
\footnotetext[1]{From Ref. \onlinecite{Moshopoulou02} for CeRhIn$_5$, repeated from
Table \ref{in_tbl}}
\footnotetext[2]{Constrained to Hg(1)-Ce}
\label{hg_tbl}
\end{table*}

\begin{figure}[t]
\includegraphics[width=3.5in]{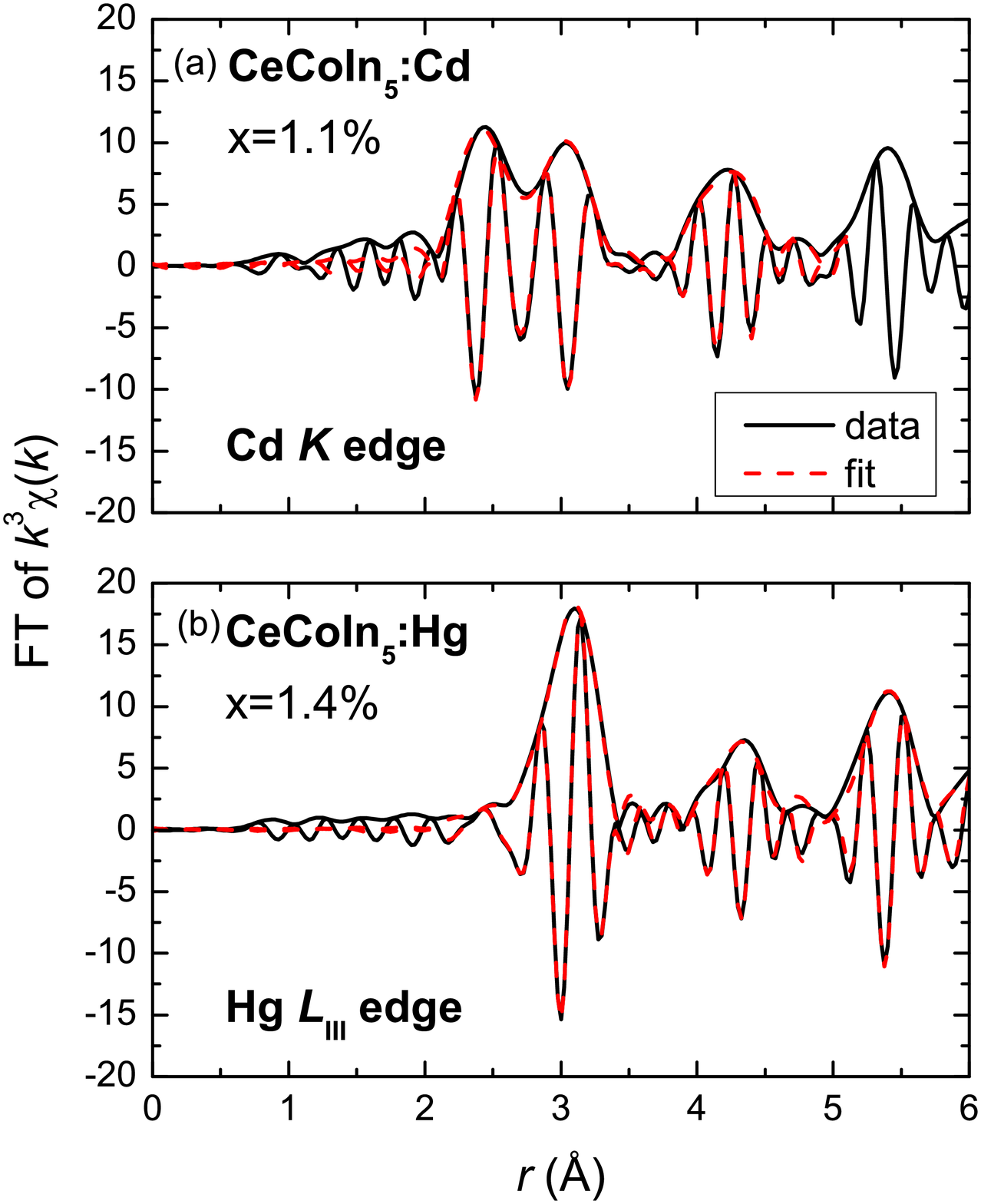}
\caption{(Color online) FT of $k^3(\chi(k)$ data (solid) and fit (dashed)
for (a) 1.1\% Cd-substituted
and (b) 1.4\% Hg-substituted CeCoIn$_5$. Note the large reduction
in the Hg(2)-Co scattering compared to the Cd(2)-Co scattering near 2.4~\AA.
}
\label{cdhg_fig}
\end{figure}

The Cd $K$-edge fit results for three of the CeCo(In$_{1-x}$Cd$_x$)$_5$ samples 
are summarized in Table \ref{cd_tbl}, and an example of the fit for $x$=0.011 is
shown in Fig. \ref{cdhg_fig}a.  Two extra constraints were necessary on the bond 
lengths due to the distortion discussed below, and the maximum $r$ in the fit range 
was limited to 5.1 \AA{} in order to reduce the effect of multiple scattering on 
determining this distortion.
The fit model describes the data very well, with 
$f_\textrm{In(1)}=0.47(4)$ for the $x$=0.011 sample. No obvious trend in 
$f_\textrm{In(1)}$ is observed with $x$, and a value of
$f_\textrm{In(1)}=0.43(3)$ describes the fits to all the Cd-substituted samples.
One can get
a rough estimate of the number of Cd on In(2) sites by comparing
the FT data in Fig. \ref{cdhg_fig}a to the In $K$-edge data in 
Fig. \ref{inrs_fig}a. These data show a reduction 
in the amplitude of the peak near
$r\sim2.5$ \AA{} of $\sim$80\% compared to the same peak from the In edge, indicating 
$0.8 \times 4/5 \approx 64$\% of the Cd sit on In(2) sites, while the 
remaining 36\% occupy 
In(1) sites, in rough agreement with the fits. The most obvious difference, 
however, is in the amplitude of the
peaks at longer pair distances. Although these can be fit by including lattice disorder via 
enhanced $\sigma^2$ parameters with a fit quality factor $R(\%)$ of about 18.4\%, the fit quality 
is substantially improved by 
allowing for a local contraction of about 0.2 \AA{} of the $c$ axis near Cd 
atoms. The $c$ axis in the fit in Table \ref{cd_tbl} is 7.32(3) \AA, compared
to  a value of 7.5513 \AA{} obtained by diffraction on pure CeCoIn$_5$.\cite{Moshopoulou02} 
Meanwhile, these Cd $K$ edge fits indicate 
$a=4.602(7)$ \AA{} and $z=0.297(7)$ (position along $c$ of In(2) plane),
in reasonable agreement with the values from diffraction on 
CeCoIn$_5$ of $a_\textrm{diff}=4.6129$ \AA{} and $z_\textrm{diff}=0.3094$.
As a consequence of this $c$ axis distortion, the overlapping Cd(1)-Ce, 
Cd(1)-In(2), In(2)-Ce, etc., peak positions are split by $\sim0.1$ \AA,
causing the dominant peak in the In edge FTs in Fig. \ref{inrs_fig}a to be 
strongly suppressed in the Cd edge FTs in Fig. \ref{cdhg_fig}. The same
argument holds for the peaks for longer pair distances.

\begin{figure}[t]
\includegraphics[width=3.5in]{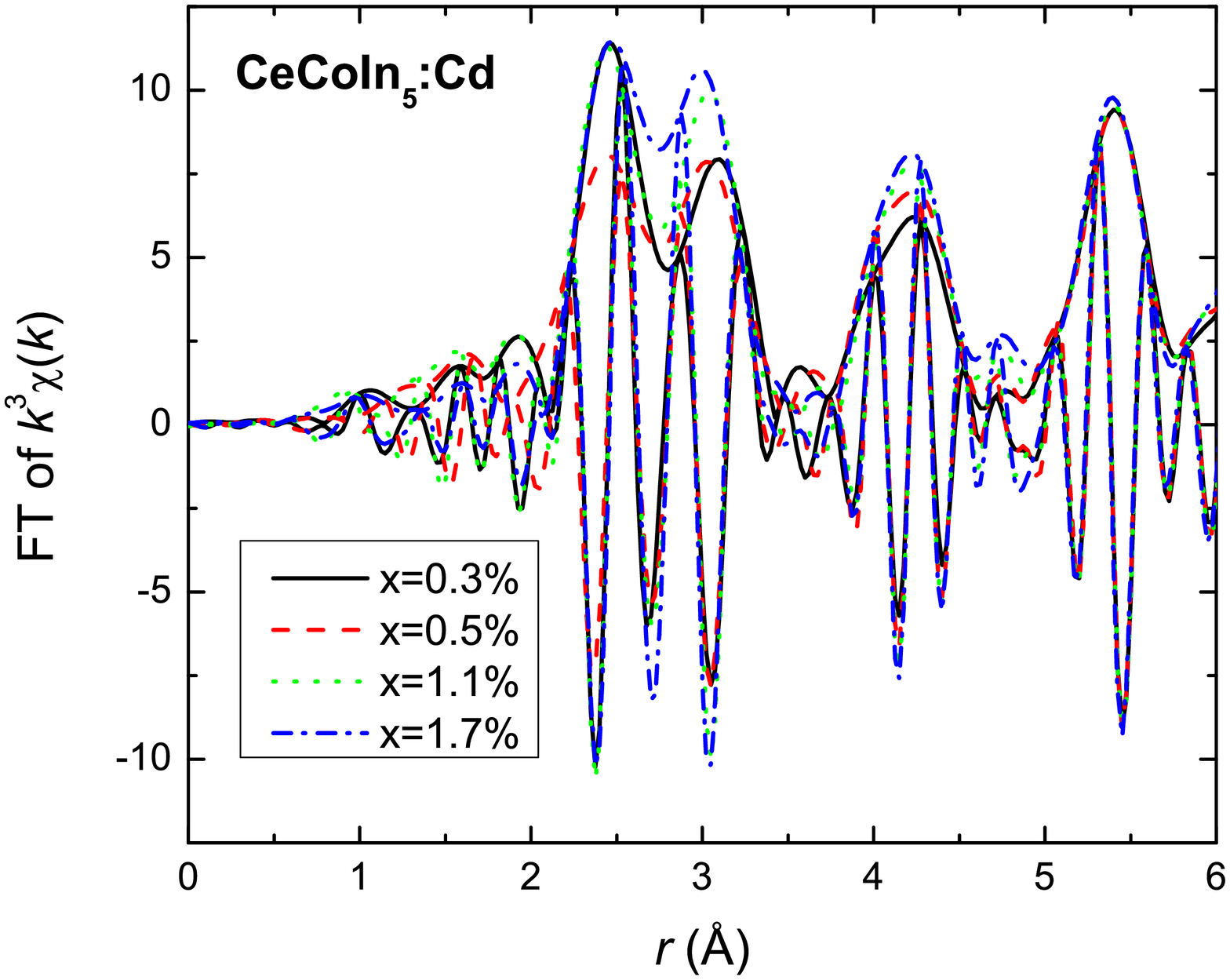}
\caption{(Color online) FT of $k^3(\chi(k)$ data for all measured Cd-substituted
samples. Note changes in local structure. Despite these changes, no clear
trends with $x$ are deduced from the fit results.
}
\label{cdcomp_fig}
\end{figure}

Data and fit results to the other Cd-substituted samples are similar, in spite of the
obvious, and apparently systematic, differences in the transforms shown in 
Fig. \ref{cdcomp_fig}. These differences are described well by the fit
parameters shown in Table \ref{cd_tbl} as being mostly due to differences in
the mean-squared distribution widths, $\sigma^2$, for the various peaks. In 
particular, no trends are observed in $f_\textrm{In(1)}$.

\begin{figure}[t]
\includegraphics[width=3.5in]{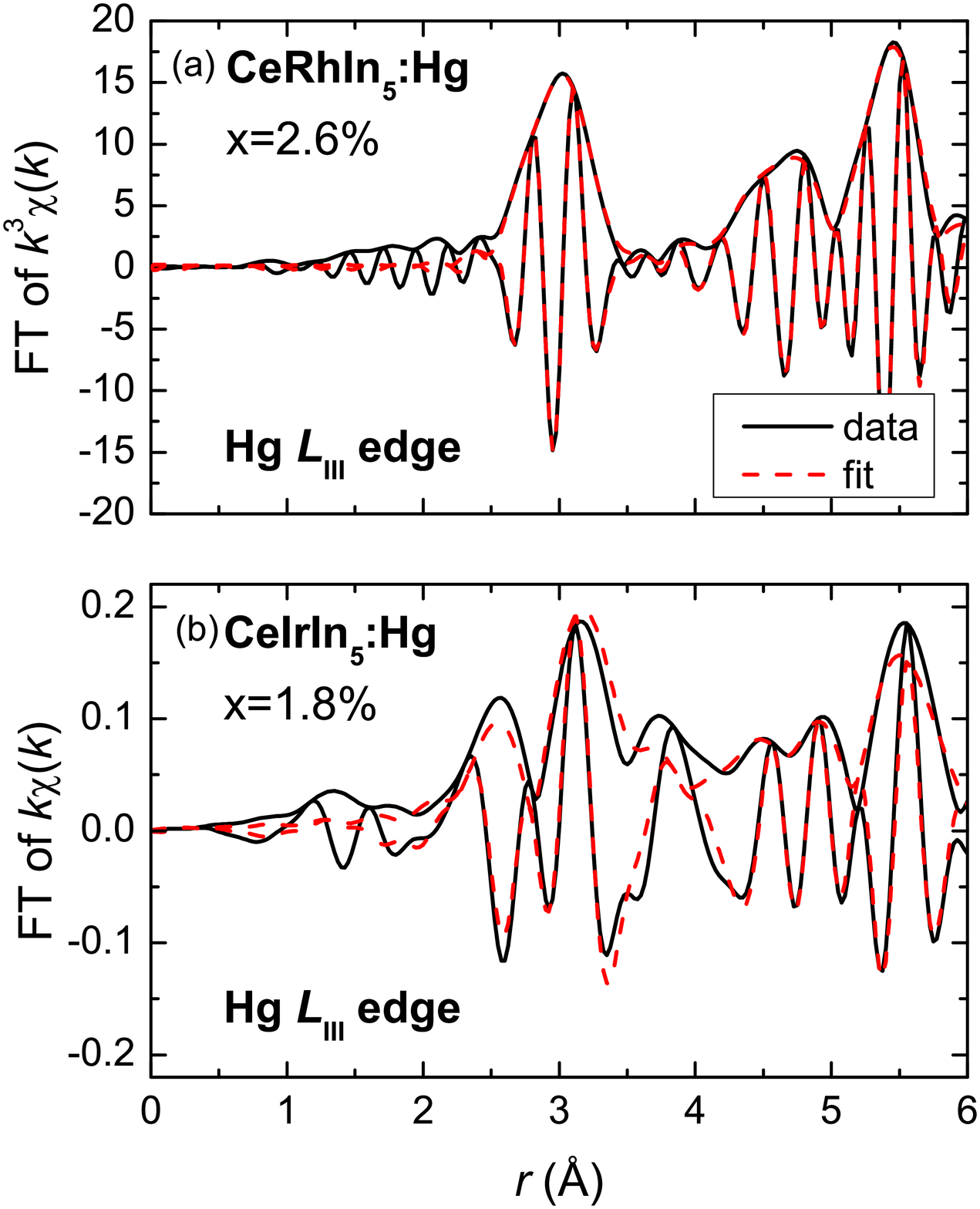}
\caption{(Color online) FT of $k^3(\chi(k)$ data (solid) and fit (dashed)
for (a) 2.6\% Hg-substituted CeRhIn$_5$
and (b) 1.8\% Hg-substituted CeIrIn$_5$. Data in panel (b) are transformed
between 2.5 and 11.5 \AA$^{-1}$ after Gaussian narrowing by 0.3 \AA$^{-1}$, 
in contrast to the 2.5 and 16.0 \AA$^{-1}$ range used for all other data 
presented here.  The apparent peak at 2.6 \AA{} is actually an interference
dip at $\sim$2.8 \AA{} as a consequence of the different transform range. No
evidence for Hg(2)-Ir scattering is observed.}
\label{rhir_fig}
\end{figure}

By comparison, the Hg-edge data are much more straightforward. Like both
the Cd- and Sn-substituted CeCoIn$_5$ samples, Hg substitutes more strongly 
onto the In(1) site than would be described by a random occupancy (20\%). 
However, Hg prefers the In(1) site even more than the other substituents, with 
$f_\textrm{In(1)}\approx70\%$ (see Table \ref{hg_tbl}). For $T$=Rh, Hg sits 
almost uniformly on the In(1) site, with little change to the local lattice. 
This result is clearly visible both in the data and fits shown in 
Figs. \ref{cdhg_fig}b and \ref{rhir_fig}a, and in the fit results in 
Table \ref{hg_tbl}.
A strong correlation exists in the fits between the $\sigma^2$ parameters from
the Hg(2) sites and
$f_\textrm{In(1)}$, whereby a large Hg(2) $\sigma^2$ reduces $f_\textrm{In(1)}$.
Such a correlation is expected for high $f_\textrm{In(1)}$, since very little,
if any, of the EXAFS signal will be coming from the Hg(2) sites, and EXAFS
amplitudes vary as 1/$\sigma$. Some of these Hg(2) $\sigma^2$ parameters had
to be limited to 0.04 \AA$^2$ in the fits to keep them from being arbitrarily large.
Data and fit results on the other CeRh(In$_{1-x}$Hg$_x$)$_5$ samples are similar,
with $f_\textrm{In(1)}=0.92(4)$.

Fit results on CeIr(In$_{1-x}$Hg$_x$)$_5$, while consistent with 
100\% of the Hg on In(1) sites, are of lesser quality (Fig. \ref{rhir_fig}b) with 
a much larger $R(\%)$ value (Table \ref{hg_tbl}), possibly indicating that
not all of the Hg substitutes into the CeIrIn$_5$ lattice.  Reducing the 
emphasis on the high-$k$ data by $k$-weighting the data, as opposed to 
$k^3$-weighting the data as done elsewhere in this study, improved
the fit somewhat, consistent with the presence of an impurity phase.
Preliminary nuclear quadrupole resonance (NQR) data also 
show that not all of the Hg is in a simple In-like site in the crystal 
lattice.\cite{Urbano_priv08}
 A strong possibility is that a small fraction of the Hg exists in another phase,
probably some kind of Hg-In binary alloy, although including scattering
paths from common Hg-In alloys, such as HgIn, did not improve the fit quality. 
It is important to recall that EXAFS selects the Hg
atoms, even though they only exist in about 1\% of the lattice, and so a 
possible 20\% Hg-phase fraction would only translate to 0.2\% of the sample,
yet still account for the misfit in the Hg-edge data. In any case, there remains
no evidence of a Hg-Ir peak near 2.8 \AA, indicating none of the Hg sits on the In(2)
site.

\section{Discussion}
\label{Discussion}

The difference in the local environment around the In(1) and In(2) sites is
substantial, with a nearest-neighbor pair distance of half an angstrom
shorter from the In(2) site. It is therefore not surprising that a given
substituent onto the In sites would prefer the In(1) site, and, in fact, the measured
distributions track the atomic radii both of the substituent species $M$ and
of the transition metal species $T$. Specifically, the calculated radii for
In is 2.00 \AA, while for Cd, Sn, and Hg, it is 1.71 \AA, 1.72 \AA, and 1.76 \AA, 
respectively.\cite{Clementi67} These values track the respective occupancies 
$f_\textrm{In(1)}$ in CeCo(In$_{1-x}M_x$)$_5$ of 43(3)\%, 55(5)\%,\cite{Daniel05b} 
and 71(5)\%, assuming no dependence on $x$.
The occupancies $f_\textrm{In(1)}$ in Ce$T$(In$_{1-x}$Hg$_x$)$_5$
also track how constricted the In(2) environment is by the $T$ species: the
atomic radii of Co, Rh, and Ir, are 1.67, 1.83, and 1.87 \AA, respectively, a situation that is also reflected in the measured In(2)-$T$ distance in the average crystal 
structures of Ce$T$In$_5$.\cite{Moshopoulou02} It is worth pointing out that
this situation is not unique in anisotropic crystal structures with two very
different sites for a given atomic species. For instance, there are two Cu
sites in YBa$_2$Cu$_3$O$_7$, and substitutions of Cu with Co are
almost uniformly on the chain Cu(1) site ({\it eg}. see Refs. \onlinecite{Miceli88}
and \onlinecite{Bridges89}).

Placing the impurity preferentially into the Ce-In(1) plane 
undoubtedly affects the progression of SC and AFM phases with $M$ concentration,
$x$. This point has been argued with respect to Sn substitutions, where it was 
found that $T_c \rightarrow 0$ K roughly when the mean separation between 
impurities within a plane is about equal to the superconducting coherence 
length in the pure material.\cite{Daniel05b} Although this may be the dominant 
effect in rapidly reducing $T_c$ with respect to $x$, more subtle effects 
likely determine the variation between samples with Cd, Sn, and Hg substitutions in 
CeCoIn$_5$. For one, 
even though $f_\textrm{In(1)}$ is slightly smaller for Cd substitutions 
compared to Sn, superconductivity is destroyed more quickly with $x$ for Cd 
than Sn (Fig. \ref{phase}). One can argue that this difference is due to the 
fundamental Ce/In charge interaction differences between these materials, since 
one is hole doped while the other is electron doped. In that case, one should 
directly compare the hole-doped, Cd- and Hg-substituted systems. The ratio 
between the critical concentrations, $x_c$, where superconductivity is 
destroyed between Cd (1.7\%) and Hg (1.4\%) is about 1.2.  This value is close 
to the square of the ratio of $f_\textrm{In(1)}$ for Hg and Cd of about 1.3, 
further supporting the notion of strong scattering for the in-plane In(1)-site
impurities.  Although this argument seems to explain differences in $x_c$ based 
on $f_\textrm{In(1)}$, it doesn't explain all the differences between Cd and 
Hg substitutions in CeCoIn$_5$. For example, $T_c$ is higher for all $0.8<x<1.4\%$ in 
Hg-substituted samples compared to Cd-substituted samples, despite the much 
higher In(1) occupancy of Hg. These results indicate that while qualitatively 
the degree of In(1)-site occupancy plays a role, the detailed electronic 
structure around an impurity is at least as important in determining 
quantitative behavior and the possible role of a ``local pressure" effect around the $M$ atom.  

The effect of structural disorder on the electronic and 
magnetic properties introduced by percent-level substitutions on the In sites 
remains enigmatic. The central dichotomy is between the observations of 
dramatic changes in the ground-state properties and the small changes in the 
lattice parameters.  In fact, one expects less than 0.004 \AA{} reduction in the
lattice parameters at $x_c$ based on the atomic radii. 
Diffraction measurements (Sec \ref{Experimental}) on the Cd-substituted material indicate
a lattice contraction consistent with this value. 
Measurements on the Sn- and Hg-substituted samples, however, have not
been able to identify such a contraction. 
In any case, such a small distortion should have a relatively small effect on the magnetic 
coupling strength $\mathcal{J}\varrho$, where $\mathcal{J}$ is the local 
moment/conduction electron exchange parameter and $\varrho$ is the electronic 
density of states at the Fermi level. For instance, 
$\mathcal{J}\propto V_{fc}^2$, the hybridization matrix element, which goes as 
$1/R^{5}$ in a tight-binding model,\cite{Harrison87,Harrison99} where $R$ is 
the distance between the Ce and In atoms. This formula implies a less than 1\% 
increase in $V_{fc}$. Countering this change, the $p$-electron orbital radius 
of all the substituents discussed here causes an overall \textit{decrease} in 
$V_{fc}$ with $x$ relative to the pure compound. Similarly, even if each 
substituent changes the local density of states by 50\%, the average change 
would be less than 1\% at $x_c$, positive for Sn and negative for Cd and Hg. 
Consequently, $\mathcal{J}$ should be nearly constant for Sn and decrease by 
less than 2\% for Cd and Hg. Because of these small changes in the average 
structure and conduction electron concentrations, it is difficult to rationalize
the dramatic changes in the ground state in these materials, even if one argues 
that the undoped system lies near a sharp band. One can argue\cite{Daniel05b} 
that the SC state is very sensitive to small amounts of disorder; however, such 
an argument can't easily apply to the sharp increase in $T_\textrm{N}$ 
observed in the Cd and Hg substituted samples.  In any case, there are other
indications that the electronic structure is remarkably sensitive to small 
local structure changes.  For example, Dynamical Mean Field 
Theory calculations on CeIrIn$_5$ indicate that the hybridization of the 
Ce-In(2) bond is stronger than that of the Ce-In(1) bond despite nearly equal 
bond distances, giving rise to two hybridization gaps in the optical 
conductivity at $\sim$30 meV and $\sim$70 meV, in agreement with 
experiment.\cite{Shim07b}  

A clue for resolving this issue of how the In-site substituents control the 
physics of the Ce$T$In$_5$ compounds comes from recent NQR experiments on 
Cd-substituted CeCoIn$_5$.\cite{Urbano07}   The In NQR data taken on pure 
CeCoIn$_5$ (SC only), 1\% (coexistent AFM and SC order), and 1.5\% (only AFM) 
Cd-substituted samples indicate that the changes in electronic structure occur 
locally around a substituent atom.  This conjecture is supported by NQR data in 
the normal state, which show that the spin-lattice relaxation rate is nearly 
identical despite radical changes in the ground state.  It is expected that 
large changes in the spin-fluctuation spectrum, and hence $1/T_1$, should occur 
in the evolution from a SC to an AFM state, and is  observed in systems such as 
CeCu$_2$(Si,Ge)$_2$.\cite{Kitaoka02}  However, if Cd nucleates magnetism on a 
scale less than the magnetic correlation length, for instance, by changing the 
Ruderman-Kittel-Kasuya-Yosida (RKKY) interaction through a change in $\varrho$ 
or $\mathcal{J}$, then there will be little change in $1/T_1$. Only when the 
substituent concentration is large enough such that the magnetic correlation 
lengths overlap does long-range order develop. For the sake of argument,
consider only correlations in the $ab$ plane and Cd atoms on the In(1).
Long-range antiferromagnetic order develops at about $x=0.8$\%, or about 
0.5\% of the In(1) sites occupied with Cd. The mean separation between Cd 
atoms along the $a$ or $b$ directions is therefore about 14 lattice spacings.
Inelastic neutron scattering measurements on CeCoIn$_5$ reveal\cite{Stock08} that the dynamic 
correlation length in the $ab$ plane is about $\xi_{ab}$ = 9.6 \AA, which is 
only about 2 lattice spacings. The antiferromagnetic droplets would then have 
to increase 7-fold in this simplified two dimensional model to overlap and 
generate long-range magnetic order.  Nuclear Magnetic Resonance measurements, 
in fact, show indications of such an increase in $\xi$ below 10 K in a
Cd-substituted sample.\cite{Curro09}

The picture that is emerging is reminiscent of the Kondo 
disorder\cite{Miranda97b,Miranda01} and antiferromagnetic Griffiths' 
phase\cite{Castro-Neto00} discussions around compounds like UCu$_4$Pd and 
U$_{1-x}$Y$_x$Pd$_3$.\cite{Stewart01} These arguments revolve around the magnetic 
coupling strength $\mathcal{J} \varrho$, and the Doniach argument regarding the 
competing Kondo interaction and RKKY effects.\cite{Doniach77}
Here, the reduction of $T_c$ in the electron-doped material occurs due to the
distribution of scattering centers and the strong 
Abrikosov-Gorkov-like (AG) scattering mechanism,\cite{Abrikosov61} only requiring
local increases in the Kondo temperature around a scattering center.\cite{Daniel05b}  
In the hole-doped, Cd- and Hg-substituted systems, $\varrho$ changes
with $x$, apparently enough to allow RKKY interactions to dominate 
over the Kondo effect, potentially allowing antiferromagnetic droplets to form within a 
Griffiths' mechanism around impurity sites, consistent with the NQR
observations.\cite{Urbano07} Within this picture, in both the electron-doped, 
Kondo-disorder/AG, regime and the hole-doped, AF Griffiths phase regime, lattice 
disorder plays a key role in the development of various properties with $x$ by allowing 
the precipitation of larger-scale perturbations, either by disturbing the coherence of 
the large SC state or by precipitating long-range magnetic 
interactions. Although these qualitative ideas may indeed play a defining role
in determining the properties in substituted 115s, a quantitative 
theory has not yet been developed that properly accounts for the details of this
quantum critical system.

\section{Conclusion}
\label{Conclusions}

The fraction of $M$ atoms on In(1) sites is determined in Ce$T$(In$_{1-x}M_x$)$_5$ as a function of $M$ in CeCo(In$_{1-x}M_x$)$_5$ and as
a function of $T$ in Ce$T$(In$_{1-x}M_x$)$_5$ with $T$ = Co, Rh, and Ir, using EXAFS
measurements at the In $K$, Cd $K$, and Hg $L_\textrm{III}$ edges. Fits to the
In $K$-edge data indicate no measurable change in the average structure with these
substituents. Fits to the Cd $K$-edge data for CeCo(In$_{1-x}$Cd$_x$)$_5$ indicate 
about $f_\textrm{In(1)}$=43(3)\% of Cd atoms 
reside on In(1) sites, independent of $x$ and similar to previous results\cite{Daniel05b} of 
$f_\textrm{In(1)}$=55(5)\% for Sn in CeCo(In$_{1-x}$Sn$_x$)$_5$. In
addition to this strong preference to occupy the In(1) site (random
occupation would be $f_\textrm{In(1)}$=20\% in this structure), the local lattice is distorted around Cd sites, consistent 
with a local decrease in the $c$ axis of about 0.2 \AA, while the $a$ lattice 
constant and the $z$ parameter describing the position of the 
In(2) planes remain unchanged. These results contrast with those from the Hg
$L_\textrm{III}$-edge data that indicate $f_\textrm{In(1)}$=71(5)\% in
CeCo(In$_{1-x}$Hg$_x$)$_5$, with
only minimal changes to the local lattice structure. Moreover, $f_\textrm{In(1)}$
increases to 92(4)\% for $T$=Rh and 100(10)\% for $T$=Ir.
While these results are
rationalized in terms of the atomic radii of the $M$ and $T$ species and gross changes
in the superconducting transition temperature, the dramatic changes in the ground 
state, especially in the hole-doped materials, are difficult to understand in terms of 
localized impurity scatterers.  Rather, a sharper division can be made based on 
electron- versus hole-doped samples and allowing for the possibility of 
antiferromagnetic droplet formation. Therefore, while strong conduction electron 
scattering around In(1)-site defects undoubtedly plays
a large, and possibly majority, role in the progression of $T_c$ with $x$,
a complete understanding of the differences in the ground states requires a more thorough understanding of the actual 
electronic structure around defect atoms and their effect on the system as a 
whole.

\section*{Acknowledgments}

Supported
by the 
U.S. Department of Energy (DOE) under Contract No. DE-AC02-05CH11231 (Lawrence
Berkeley Natonal Laboratory), by the National Science Foundation (NSF) under 
grant DMR 0854781 (University of California, Irvine) and grant DMR-0756281
(Louisina State Univerisity), and further supported by the U.S. DOE 
(Los Alamos).
X-ray absorption data were collected at the Stanford Synchrotron Radiation 
Lightsource, a national user facility operated by Stanford University on
behalf of the DOE, Office of Basic Energy Sciences. 

\bibliography{/home/hahn/chbooth/papers/bib/bibli}

\end{document}